\documentclass[fleqn,usenatbib]{mnras}
\usepackage{newtxtext,newtxmath}
\usepackage[T1]{fontenc}
\DeclareRobustCommand{\VAN}[3]{#2}
\let\VANthebibliography\thebibliography
\def\thebibliography{\DeclareRobustCommand{\VAN}[3]{##3}\VANthebibliography}
\usepackage{graphicx}
\usepackage{amsmath}
\usepackage{mathrsfs}
\usepackage{bm}
\usepackage{xcolor}
\usepackage{soul}
\usepackage{url}
\usepackage{xfrac}
\usepackage{tabularx}
\usepackage{float}
\usepackage{ulem}

\title[vMF Plutino Orbital Pole Distribution]{A von Mises--Fisher Distribution for the Orbital Poles of the Plutinos}
\author[I. C. Matheson, R. Malhotra, J. T. Keane]{
Ian C. Matheson,$^{1}$\thanks{Email: ianmatheson@arizona.edu}
Renu Malhotra,$^{2}$\thanks{Email: malhotra@arizona.edu}
James T. Keane$^{3}$\thanks{Email: james.t.keane@jpl.nasa.gov}
\\
$^{1}$Department of Aerospace \& Mechanical Engineering, The University of Arizona, 1130 N. Mountain Ave., P.O. Box 210119, Tucson, AZ 85721, USA\\
$^{2}$Lunar \& Planetary Laboratory, The University of Arizona, 1629 E. University Blvd., P.O. Box 210092, Tucson, AZ 85721, USA\\
$^{3}$Jet Propulsion Laboratory, California Institute of Technology, 4800 Oak Grove Dr., Pasadena, CA 91109, USA
}

\date{Accepted 2023 April 18. Received 2023 April 11; in original form 2022 October 27.}

\pubyear{2023}

\begin{document}
\label{firstpage}
\pagerange{\pageref{firstpage}--\pageref{lastpage}}
\maketitle

\begin{abstract}
\noindent
Small solar system bodies have widely dispersed orbital poles, posing challenges to dynamical models of solar system origin and evolution.  To characterize the orbit pole distribution of dynamical groups of small bodies it helps to have a functional form for a model of the distribution function.  Previous studies have used the small-inclination approximation and adopted variations of the normal distribution to model orbital inclination dispersions.  Because the orbital pole is a directional variable, its distribution can be more appropriately modeled with directional statistics.  We describe the von Mises--Fisher (vMF) distribution on the surface of the unit sphere for application to small bodies' orbital poles.  We apply it to the orbit pole distribution of the observed Plutinos.  We find a mean pole located at inclination $i_0=3.57^\circ$ and longitude of ascending node $\Omega_0=124.38^\circ$ (in the J2000 reference frame), with a 99.7 per cent confidence cone of half-angle $1.68^\circ$.  We also estimate a debiased mean pole located $4.6^\circ$ away, at $i_0=2.26^\circ, \Omega_0=292.69^\circ$, of similar-size confidence cone.  The vMF concentration parameter of Plutino inclinations (relative to either mean pole estimate) is $\kappa=31.6$.  This resembles a Rayleigh distribution function, with width parameter $\sigma=10.2^\circ$.  Unlike previous models, the vMF model naturally accommodates all physical inclinations (and no others), whereas Rayleigh or Gaussian models must be truncated to the physical inclination range 0--180$^\circ$. Further work is needed to produce a theory for the mean pole of the Plutinos against which to compare the observational results.
\end{abstract}


\section{Introduction}
\label{s:introduction}

The orbital distribution of small bodies in the solar system has been of great interest for its value in providing insights into the origin and evolution of our solar system~\citep{Nesvorny:2018}.
One of the observations about solar system small bodies that has been an abiding challenge to theoretical models is the wide dispersion of orbital inclinations found in both the asteroid belt and the Kuiper belt.
For example, asteroids in the main asteroid belt are found with orbital poles dispersed by over $60^\circ$ relative to the orientation of the solar system's total angular momentum vector and the orbital poles of Kuiper belt objects range over $140^\circ$.
In contrast, the major planets of the solar system have orbital poles within a few degrees of the ecliptic pole.

Proposed models of the dynamical history of the solar system often look to testable predictions for the inclination distribution of the small bodies, especially for dynamical subgroups of small bodies, such as the Jupiter Trojan asteroids~\citep{Levison:2021} and the resonant Kuiper belt objects \citep[e.g.][]{Nesvorny:2016}.
It is therefore useful to have quantitative models to characterize the distribution of orbital inclinations of the small bodies, including dynamical subgroups of particular interest.

For solar system orbits, the orientation of an orbital plane is described by the directional unit vector along the orbit pole,
\begin{equation}
    \hat{\bm{h}} = (h_x,h_y,h_z) = (\sin{i}\sin{\Omega}, -\sin{i}\cos{\Omega}, \cos{i})
\end{equation}
where $i,\Omega$ are the orbital inclination and the longitude of ascending node, respectively, in the J2000 reference frame.
(In this reference frame, the zero-inclination unit vector $\hat{\bm{k}}=(0,0,1)$ is the pole of the ecliptic at the epoch January 1, 2000.) The orbit poles of a group of small bodies will then appear as a distribution of points on the surface of the unit sphere.

In theoretical analyses in solar system dynamics, it is common to limit to small inclinations (relative to the ecliptic), make the approximation $\sin i\approx i$, and to adopt the two-component "inclination vector" defined as
\begin{equation}
    (q,p) = (i\cos{\Omega},i\sin{\Omega}).
\label{e:qp}
\end{equation}
For example, Laplace-Lagrange linear secular theory describes, to a good approximation, the time evolution of the two-component inclination vector of non-resonant test particles as a circulation
around a "forced" orbit pole given by
\begin{equation}
(q_0,p_0)=(i_0\cos{\Omega_0},i_0\sin{\Omega_0})
\label{e:q0p0}
\end{equation}
where the forced inclination $i_0$ and the forced longitude of node $\Omega_0$ are determined by the particle's semi-major axis and by the secular perturbations of the planets, with all $q$'s and $p$'s, including those of the giant planets, varying with a superposition of sinusoidal variations on secular timescales \citep{md99,cc08}.
Consequently, measurement of the mean pole of a population of small bodies can serve to test dynamical effects of planetary perturbations \citep[e.g.][]{vm17,ossos14}, and the distribution of their inclinations can serve to test theoretical models of the dynamical history of the planetary system \citep[e.g.][]{Nesvorny:2015,Volk:2019}.

Because the orbit pole is a directional variable, it is more appropriate to use the tools of directional statistics for quantitative modeling of the orbit pole distributions of small bodies.
However, previous studies have generally been limited to modeling the scalar inclination variable, $i$.
For example, \citet{brown_2001} adopted a functional form for the inclination distribution of Kuiper belt objects as proportional to the sine of the inclination, multiplied by a zero-mean Gaussian distribution, for inclinations measured relative to the ecliptic:
\begin{equation}
f(i)\propto\sin{i}\exp{\frac{-i^2}{2\sigma^2}}.
\label{e:fiB01}
\end{equation}
This form was subsequently adopted by a number of other authors \citep[e.g.][]{gulbis2010unbiased,Nesvorny:2012,vm17,ossos14}.
For low inclinations, Eq.~\ref{e:fiB01} simplifies to a Rayleigh distribution function for $i$,
\begin{equation}
f(i)\propto i\exp{\frac{-i^2}{2\sigma^2}}.
\label{e:fiRi}
\end{equation}
When $i$ and $\Omega$ are considered jointly as in equation (\ref{e:qp}), this becomes a bivariate Gaussian distribution in $q$ and $p$ with zero mean in each component, zero corelation between the two components, and with the standard deviation in each component equal to $\sigma$.

If the inclinations are taken to have a Rayleigh distribution and the longitude of ascending node is taken to have a uniform distribution relative to a reference plane distinct from the ecliptic plane, then
\begin{equation}
(q,p)=(q_0,p_0)+(q_1,p_1)
\end{equation}
can be treated as a bivariate Gaussian distribution centred around $(q_0,p_0)$, where $(q_0,p_0)$ is the "forced" orbit pole and $(q_1,p_1)$ is the "free" component.
Then, $i$ will no longer be Rayleigh distributed with respect to the ecliptic, and $\Omega$ will no longer be uniformly distributed with respect to the ecliptic.

As the inclination dispersion of a small body population increases, or as the mean plane of the population increases in inclination, then Eq.~\ref{e:fiB01} or Eq.~\ref{e:fiRi} or the equivalent bivariate Gaussian distribution in the $(q,p)$ plane, all become less useful as a functional form to describe the distribution of the orbit poles of the population for at least two reasons.
First, to model a sample with a large inclination dispersion by using a Rayleigh distribution (or the bivariate Gaussian distribution) it becomes necessary to truncate the distribution function to the range $0\le i \le \pi$ radians to abide by the physical limits, as in \citet{lin2019evidence}.
Second, the Rayleigh distribution for $i$ (or the equivalent bivariate Gaussian distribution for $q,p$) depends on the small-angle approximation $\sin{i}\approx i$.
At high inclinations above 30--45$^\circ$, this approximation holds to only one significant digit.
It is then more useful to consider the full orbit pole unit vector $\hat{\bm{h}}$ without projecting on the ecliptic and without the low-inclination approximation.
A distribution of orbit poles is represented by a distribution of points on the unit sphere.

We propose that the von Mises--Fisher (vMF) distribution is more suitable to quantify the distribution of orbit poles of dynamical groups of small bodies having high inclination dispersions.
The vMF distribution on the surface of the unit sphere is the analog of the isotropic bivariate Gaussian distribution.
It has circular contours of constant probability density centred about a mean direction, and its concentration parameter, $\kappa$, is straightforwardly related to the analogous parameter $\sigma$ for the Gaussian distribution.
The vMF distribution is part of a larger class of distribution functions useful for modeling directional data.
There exists a significant amount of literature on directional statistics; we refer the interested reader to the books by \citet{Jupp:2009} and \citet{fisher_1993}.
Directional statistics has been more commonly applied in meteorology and geology and bioinformatics.
The orbit plane distribution of small bodies in the solar system is a natural application as well.

In this paper, we first describe the vMF distribution, then, for illustration, we apply it to the observed population of Plutinos to measure the dispersion of their orbital poles relative to their mean pole.
We choose the Plutinos as an illustrative example for three reasons.
First, the Plutinos are a well-defined and dynamically important population with a statistically significant sample size of 431 (see Section \ref{s:Plutinos}).
Second, over 30 per cent of this sample have orbital inclinations exceeding 15$^\circ$ relative to the ecliptic.
Third, we are intrigued by this population because there is no theoretical prediction in the literature for its mean orbit pole.
This is in contrast with non-resonant populations of small bodies for which the Laplace-Lagrange secular theory provides a prediction of the mean orbit pole forced by the major planets.

The rest of this paper is organized as follows.
\begin{itemize}
\item In Section~\ref{s:vmf_distribution} we describe the vMF distribution and how to estimate its parameters from a sample of unit vectors.
We also describe how the vMF distribution is related to the univariate Rayleigh distribution; the latter is equivalent to the isotropic bivariate Gaussian distribution in in the limit of a highly concentrated sample.
\item In Section~\ref{s:vm17_method} we describe an alternate way to estimate the mean pole of a sample of Plutinos, for comparison.
\item In Section~\ref{s:Plutinos}, we identify our sample of Plutinos.
\item In Section~\ref{s:plutinos_results}, we conduct the vMF parameter estimation and goodness-of-fit tests described in Section~\ref{s:vmf_distribution} for the Plutino sample, and also compute the alternate mean pole described in Section~\ref{s:vm17_method}.
\item In Section~\ref{s:discussion}, we discuss the results and suggest future directions for improving quantitative modeling of the distributions of orbit poles of small bodies in the solar system.
\item In Appendix \ref{clones}, we discuss measures taken to account for uncertainty in the orbits of the Plutinos.
\end{itemize}

\section{von Mises--Fisher distribution} \label{s:vmf_distribution}

For a distribution of points on the surface of the unit sphere, the  analog of the isotropic bivariate Gaussian distribution is the von Mises--Fisher (vMF) distribution.
This distribution is unimodal and rotationally symmetric about its modal direction.
We fit the vMF distribution to a sample of $n$ unit vectors $\bm{x}_i=(x_i,y_i,z_i)$ as follows.

\begin{enumerate}
    \item First, we estimate the mean direction of the sample.
    \item Second, we estimate a conical confidence region for the mean direction.
    \item Third, we estimate the concentration parameter of the sample.
\end{enumerate}
In the following subsections, we define the vMF distribution and each step of the procedure just described.
We then describe how colatitudes and longitudes are measured relative to the mean direction.
Finally, we describe the relationship of the vMF distribution on the unit sphere to the more familiar Rayleigh distribution on the positive real line and the bivariate Gaussian distribution on the plane.
This discussion follows \citet{fisher_1993}, and applies to large samples of size $n\geq 25$.

\subsection{vMF distribution function}
\label{ss:vmf_distribution}

\begin{figure}
\includegraphics[width=0.7\columnwidth]{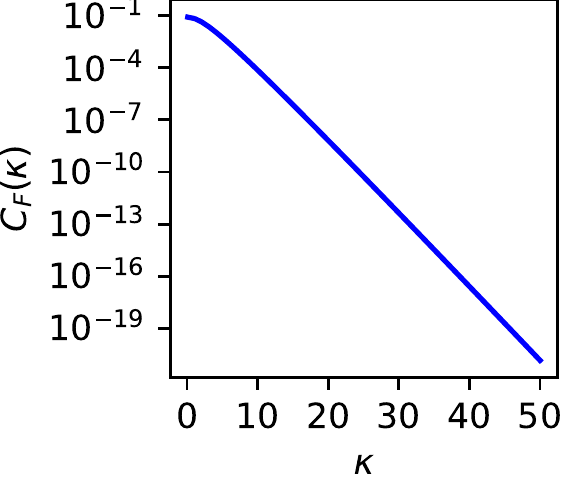}
\vglue-0.1truein\caption{The normalization parameter $C_F(\kappa)$ as found in Eq.~\ref{e:fisher_4.20}.}
\label{fig:c3_kappa}
\end{figure}

The vMF distribution is described by the following expression with domain over the surface of the sphere:
\begin{equation}
h(\theta,\phi)=C_F\exp{[\kappa(\sin{\theta}\sin{\alpha}\cos{(\phi-\beta)}+\cos{\theta}\cos{\alpha})]},
\label{e:fisher_4.19}
\end{equation}
where $\theta$ is the colatitude of a random variable and $\phi$ is its longitude, and $\alpha$ and $\beta$ are respectively the colatitude and the longitude of the mean direction of the vMF distribution, $\kappa>0$ is the non-negative concentration parameter, and the coefficient $C_F$ (plotted in Fig. \ref{fig:c3_kappa}) is defined as
\begin{equation}
C_F=\frac{\kappa}{4\pi\sinh{\kappa}}=\frac{\kappa}{2\pi(e^{\kappa}-e^{-\kappa})}
\label{e:fisher_4.20}
\end{equation}
According to this notation, a point on the surface of the sphere is defined as
\begin{equation}
(x,y,z)=(\sin{\theta}\cos{\phi},\sin{\theta}\sin{\phi},\cos{\theta}).
\label{e:fisher_pg_70}
\end{equation}
To recover the total probability mass of unity, we integrate $h(\theta,\phi)$ over the surface of the sphere as
\begin{equation}
\int_{\theta=0}^{\theta=\pi} \int_{\phi=0}^{\phi=2\pi} h(\theta,\phi)\sin{\theta}\,d\theta\,d\phi \equiv 1 .
\label{e:integrate_fisher_4.20}
\end{equation}
Note the finite range of $\theta$ (equivalently, the inclination to the reference pole direction) in this distribution, whereas previous work on orbit pole distribution functions has typically used variations of the normal distribution which has infinite domain.

\subsection{Estimating the mean direction}
\label{ss:estimate_sample_mean}

With the assumption that the sample is drawn from a unimodal and rotationally symmetric distribution, we use the sample to estimate the mean direction of the source distribution.
(Later, in Section~\ref{s:vm17_method}, we discuss a method to mitigate systematic error in the mean direction in the context of solar system small bodies' orbital poles when the data has observational selection biases.)

Given $n$ unit vectors $\bm{x}_i=(x_i,y_i,z_i)$ on the surface of the unit sphere, we compute the resultant vector $\bm{S}$ as
\begin{equation}
\bm{S}=(S_x,S_y,S_z)=\sum_{i=1}^{n} (x_i,y_i,z_i)
\label{e:fisher_3.2}
\end{equation}
and define its magnitude as
\begin{equation}
R=\|\bm{S}\|=\sqrt{S_x^2+S_y^2+S_z^2}.
\label{e:fisher_3.3}
\end{equation}
The resultant vector $\bm{S}$ points in a direction with colatitude $\hat{\theta}$ and longitude $\hat{\phi}$.
The unit vector $(\hat{x},\hat{y},\hat{z})$ in this direction is
\begin{equation}
\hat{\bm{x}}=\frac{\bm{S}}{R}=(\sin{\hat{\theta}}\cos{\hat{\phi}},\sin{\hat{\theta}}\sin{\hat{\phi}},\cos{\hat{\theta}}).
\label{e:fisher_3.4}
\end{equation}
Given the sample vectors $\bm{x}_i$, $\hat{\bm{x}}$ is the maximum-likelihood estimate of the vMF mean direction. The corresponding  maximum-likelihood estimates for the colatitude and longitude of the mean direction are respectively $\hat{\alpha}=\hat{\theta}$ and $\hat{\beta}=\hat{\phi}$.

\subsection{Estimating a confidence region for the mean direction}
\label{ss:estimate_confidence_region}

For sufficiently large sample sizes, $n\geq25$, the first two components $\hat{x}$ and $\hat{y}$ of the sample mean direction approximately follow a normal distribution.
We compute the spherical standard error, $\hat\sigma$, as follows.
We compute $\bar{R}$, and $d$,
\begin{equation}
\bar{R} = \frac{R}{n} ,
\end{equation}
\begin{equation}
d = 1-\frac{1}{n}\sum_{i=1}^n{(\bm{x}_i\cdot \hat{\bm{x}})^2}.
\label{e:fisher_5.11}
\end{equation}
Then,
\begin{equation}
\hat{\sigma}=\sqrt{\frac{d}{n\bar{R}^2}}.
\label{e:fisher_5.12}
\end{equation}
If we wish to construct a $100(1-A)$ per cent confidence region for the mean direction $(\alpha,\beta)$ centered on the sample mean direction $(\hat{\alpha},\hat{\beta})$, we define the region as the surface of the unit sphere located within a cone that has its apex at the origin, an axial direction of $(\hat{\alpha},\hat{\beta})$, and a half-angle of
\begin{equation}
q=\arcsin{(\hat{\sigma}\sqrt{-\log{A}})}.
\label{e:fisher_5.13}
\end{equation}
This approximation makes use of the same small-inclination assumption that we are trying to avoid by substituting a vMF distribution on the sphere for a bivariate planar Gaussian distribution of $(h_x,h_y)$,
but we defer a more sophisticated confidence region for later work.

\subsection{Estimating the concentration parameter}
\label{ss:estimate_kappa}

To find the maximum-likelihood estimate of the concentration parameter $\kappa$, we can numerically solve
\begin{equation}
\coth{\kappa}-\frac{1}{\kappa}=\frac{R}{n},
\label{e:fisher_5.22}
\end{equation}
or, for $\bar{R}\geq0.95$,
\begin{equation}
\hat{\kappa}\approx\frac{n-1}{n-R}.
\label{e:fisher_5.25}
\end{equation}
A $100(1-A)$ per cent confidence interval for $\kappa$ may be constructed with the respective lower and upper bounds as
\begin{equation}
\kappa_L=\frac{1}{2}\frac{\chi_{2n-2}^2{\big(1-\frac{1}{2}A\big)}}{n-R},\,\,\,\kappa_U=\frac{1}{2}\frac{\chi_{2n-2}^2{\big(\frac{1}{2}A\big)}}{n-R}.
\label{e:fisher_5.37}
\end{equation}

\subsection{Defining relative colatitude and relative longitude}
\label{ss:relative_colatitude_and_longitude}
To define the colatitude and longitude of each unit vector in the sample relative to the mean direction, we rotate the sample data so that the sample mean pole becomes the ecliptic pole $(0,0,1)$ while preserving the colatitudes and longitudes of the individual samples relative to each other.
The rotated samples $\bm{x}_i'=(\sin{\theta'}\cos{\phi'},\sin{\theta'}\sin{\phi'},\cos{\theta'})$ are related to the original samples $\bm{x}_i=(\sin{\theta}\cos{\phi},\sin{\theta}\sin{\phi},\cos{\theta})$ as
\begin{equation}
\bm{x}_i'=A(\hat{\alpha},\hat{\beta},0)\bm{x}_i,
\label{e:fisher_3.8}
\end{equation}
where
\begin{equation}
A(\hat{\alpha},\hat{\beta},0)=\begin{pmatrix}
\cos{\hat{\alpha}}\cos{\hat{\beta}} & \cos{\hat{\alpha}}\sin{\hat{\beta}} & -\sin{\hat{\alpha}} \\
-\sin{\hat{\beta}} & \cos{\hat{\beta}} & 0 \\
\sin{\hat{\alpha}}\cos{\hat{\beta}} & \sin{\hat{\alpha}}\sin{\hat{\beta}} & \cos{\hat{\alpha}}
\end{pmatrix}.
\label{e:fisher_3.9}
\end{equation}

\subsection{Relationship of the vMF distribution to the univariate Rayleigh distribution}
\label{ss:relate_vmf_to_rayleigh}

To relate the vMF distribution on the two-dimensional surface of the unit sphere to the univariate Rayleigh distribution for the colatitude relative to the mean direction, we develop an approximate expression that relates $\sigma$, the width parameter of the latter, to $\kappa$, the concentration parameter of the vMF distribution, in the limit of a highly concentrated set of random orbit poles, i.e., $\kappa\gg 1$.

The colatitude distribution relative to the mean direction $(\alpha,\beta)$ may be found by setting the mean colatitude $\alpha=0$ in Eq. \ref{e:fisher_4.19}, giving
\begin{equation}
h(\theta,\phi)=C_F\exp{(\kappa\cos{\theta})},
\label{e:fisher_4.23}
\end{equation}
where $\theta$ is now the colatitude relative to the mean direction.
From Eq. \ref{e:fisher_4.20}, the probability density function for the colatitudes and longitudes relative to the mean direction is
\begin{equation}
f(\theta,\phi)=h(\theta,\phi)\sin{\theta}=C_F\exp{(\kappa\cos{\theta})}\sin{\theta}.
\label{e:fisher_4.28}
\end{equation}
From this we get the marginal distribution of the colatitudes as
\begin{equation}
f(\theta)=\int_0^{2\pi} f(\theta,\phi)\,d\phi=2\pi\, C_F\exp{(\kappa\cos{\theta})}\sin{\theta}.
\label{e:fisher_4.29}
\end{equation}
This is
\begin{equation}
f(\theta)=\frac{\kappa}{e^\kappa-e^{-\kappa}}\exp{(\kappa\cos{\theta})}\sin{\theta}.
\label{e:fisher_4.29a}
\end{equation}

An equivalent expression is as follows:
\begin{equation}
    f(\theta) = \frac{\kappa}{1 - e^{-2\kappa}}  \sin{\theta}\,\exp{\Big(-2\kappa\sin^2{\frac{1}{2}\theta}\Big)}.
\label{e:fisher_4.29b}
\end{equation}
This form is better suited to avoid numerical inaccuracies when $\kappa$ is not too small.

In the limit of a highly concentrated sample of random unit vectors, i.e., $\kappa\gg1$ and $\theta\ll1$, $f(\theta)$ can be approximated as follows:
\begin{equation}
    f(\theta) \simeq \kappa\,\theta \,  \exp{\Big(-\frac{\kappa}{2}{\theta^2}\Big)}.
\label{e:fisher_4.29c}
\end{equation}
Now we can observe that if we take $\theta$ as the random variable then the probability density function in Eq.~\ref{e:fisher_4.29c} resembles the usual Rayleigh probability density function with width parameter $\sigma$,
\begin{equation}
    \text{Rayleigh pdf}:\  f(u;\sigma) = \frac{1}{\sigma^2}\, u\, \exp{\Big(-\frac{u^2}{2\sigma^2}\Big)}
    \label{e:rayleigh}
\end{equation}
Comparing Eq.~\ref{e:fisher_4.29c} with Eq.~\ref{e:rayleigh}, we derive the result that
\begin{equation}
\sigma\approx\frac{1}{\sqrt\kappa}.
\label{e:sigma_i}
\end{equation}

We write this as an approximate result rather than exact because the comparison is not exact: the domain of the $\theta$ distribution is limited to $[0,\pi]$ whereas the standard Rayleigh distribution has domain $[0,\infty]$.
Consequently, when applied to directional statistics, the latter must be truncated to the physical domain of colatitudes.
This introduces a correction to the normalization factor so that the truncated Rayleigh pdf is given by
\begin{equation}
 \tilde{f}(u;\sigma) = \frac{C_R (\sigma)}{\sigma^2}\,u\, \exp{\Big(- \frac{u^2}{2\sigma^2}\Big)},
    \label{e:rayleigh2}
\end{equation}
with
\begin{equation}
    C_R(\sigma) =
    \Big[1-\exp{\Big(-\frac{\pi^2}{2\sigma^2}\Big)}\Big]^{-1}  \qquad\text{for}\ 0\le u\le\pi.
\label{e:CR}
\end{equation}

To compare the vMF relative colatitude distribution in Eq.~\ref{e:fisher_4.29c} to the truncated Rayleigh distribution in Eq.~\ref{e:rayleigh2}, we also want the maximum likelihood estimate $\hat{\sigma}_\textrm{MLE}$ of the width parameter for the truncated Rayleigh distribution.
For a sample of $n$ colatitudes $u_i$, the likelihood function for $\sigma$ is
\begin{equation}
    \textrm{lik}(\sigma)=\prod_{i=1}^n \tilde{f}(u_i;\sigma)
\label{e:likelihood_function}
\end{equation}
The log-likelihood function is
\begin{equation}
\ell(\sigma)=\textrm{log}(\textrm{lik}(\sigma)),
\label{e:log_likelihood}
\end{equation}
and the maximum likelihood estimate $\hat{\sigma}_\textrm{MLE}$ is the value of $\sigma$ that solves
\begin{equation}
\frac{d\ell}{d\sigma}=-\frac{2n}{\sigma}+\frac{1}{\sigma^3}\sum_{i=1}^n u_i^2+\bigg(\frac{n\pi^2}{\sigma^3}\bigg)\big(C_R(\sigma)-1\big)=0.
\label{e:dl_dsigma}
\end{equation}
We solve Eq.~\ref{e:dl_dsigma} for $\hat{\sigma}_\textrm{MLE}$ numerically.

As an aside, we note that if we take $s=\sin\frac{1}{2}\theta$ as the random variable, then the vMF distribution for relative colatitude, $f(\theta)$ (Eq.~\ref{e:fisher_4.29a}), can be expressed as a pdf for $s$ as follows:
\begin{equation}
    f_s(s) = \frac{4\kappa}{1 - e^{-2\kappa}}\,  s \,\exp{(-2\kappa s^2)}, \quad 0\le s\le 1.
\label{e:fvMFs}
\end{equation}
In the limit of high concentration parameter, $\kappa\gg1$, this resembles a Rayleigh pdf with width parameter
\begin{equation}
    \sigma_s\approx \frac{1}{2\sqrt\kappa} .
    \label{e:sigma_s}
\end{equation}

\begin{figure}
\includegraphics[width=3.2in]{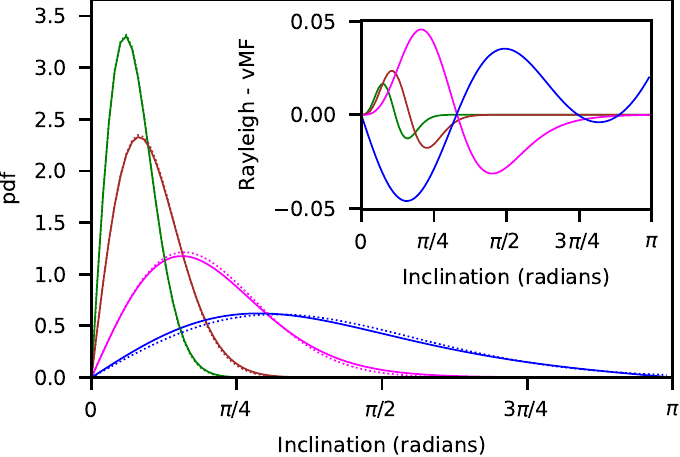}
\vglue-0.1truein\caption{The relative colatitude pdf of the vMF distribution of Eq.~\ref{e:fisher_4.29b} (solid lines) is compared with the truncated Rayleigh distribution of Eq.~\ref{e:rayleigh2} (dotted lines).
The pdfs are plotted for four values of the vMF concentration parameter, $\kappa=30,15,4,1$, in green, brown, magenta and blue, respectively.
For the corresponding Rayleigh pdf, we take $\sigma=1/\sqrt\kappa$.
The inset plots the difference between the vMF and the Rayleigh pdfs.}
\label{fig:comp}
\end{figure}

In Fig. \ref{fig:comp}, we plot the relative (marginal) colatitude pdf (Eq. \ref{e:fisher_4.29b}) of the vMF distribution along with the corresponding truncated Rayleigh distribution (Eq. \ref{e:rayleigh2}), for several values of $\kappa$.
An inset plot shows the difference between the vMF colatitude and the Rayleigh colatitude distribution functions for each concentration.
As the concentration increases, the truncated Rayleigh distribution becomes a better approximation of the exact vMF relative colatitude distribution.

With the orbital inclination identified with the colatitude in the vMF distribution, the above derivation provides the foundation for the inclination distribution function (Eq.~\ref{e:fiB01} above) adopted in \citet{brown_2001}.
It should be noted that in that paper the author made the small-$\theta$ approximation in the exponential factor of Eq. \ref{e:fisher_4.29b} but not in the $\sin{\theta}$ pre-factor, and measured the inclination $i$ (that is, $\theta$) from the ecliptic pole rather than from the mean pole of observational samples of KBOs.

\subsection{Relationship of the vMF distribution to the bivariate Gaussian distribution}
\label{ss:relate_vmf_to_gaussian}

In the high-concentration limit, $\kappa\gg1$, the vMF distribution resembles an isotropic Gaussian distribution centered around the mean direction.
Said more precisely, if we take a population of unit vectors $\bm{x}_i$ from a highly concentrated vMF distribution and rotate them as in Eq. \ref{e:fisher_3.8} such that the mean direction of the distribution is taken to the reference $z$-direction  $(0,0,1)$, then the distribution of the first two components $(x_i',y_i')$ of the rotated vectors $\bm{x}_i'$ will resemble a two-dimensional Gaussian distribution with zero correlation and equal variance in each dimension.

From Eq. \ref{e:fisher_4.28}, the marginal distribution of the longitudes in a vMF distribution, relative to the mean direction, is
\begin{equation}
f(\phi)=\int_0^\pi f(\theta,\phi)\,d\theta=\rm{constant}.
\label{e:marginal_longitudes}
\end{equation}
In other words, the relative longitudes have a uniform distribution on the circle $[0,2\pi)$.

Now consider the univariate Rayleigh pdf, which we have already shown to resemble the relative colatitude pdf of the vMF distribution for $\kappa\gg1$.

The univariate Rayleigh pdf is related to the bivariate Gaussian pdf of zero mean and zero correlation as follows. Consider the two-component random vector
\begin{equation}
    \bm{u} = (u_x,u_y) \qquad -\infty\le u_x,u_y \le\infty,
\end{equation}
whose distribution is described by the bivariate Gaussian pdf of zero mean, zero correlation and standard deviation $\sigma$,
\begin{equation}
f_2({u_x,u_y}) =
\frac{1}{2\pi\sigma^2} \exp{\Big(-\frac{u_x^2+u_y^2}{2\sigma^2}\Big)}.
\end{equation}
The zero correlation condition means that the distribution of its orientation angle is uniform random over the domain $[0,2\pi)$.
We can derive the pdf $f_u(u)$ of the length, $u$, by converting to planar polar coordinates, $(u_x,u_y) = u(\cos\phi,\sin\phi)$ and integrating the probability element $f_2({u_x,u_y})\,dx\,dy = f_2({u_x,u_y})\,u\,du\,d\phi$ over the orientation angle:
\begin{eqnarray}
f_u(u)du &=& \int_0^{2\pi} d\phi\, f_2({u_x,u_y}) \,u\,du \nonumber\\
    &=& \frac{1}{2\pi\sigma^2} \int_0^{2\pi} d\phi \,u\,du \exp{\Big(-\frac{u^2}{2\sigma^2}\Big)}\nonumber\\
    &=& \frac{u}{\sigma^2}\,\exp{\Big(-\frac{u^2}{2\sigma^2}\Big)}\, du .
\end{eqnarray}
This shows that the bivariate Gaussian distribution of zero mean, zero correlation, and equal variance in $x$ and $y$ is equivalent to the distribution of a two-component random vector that has a Rayleigh-distributed length and a uniformly-distributed orientation angle.

\section{Mitigating observational bias}
\label{s:vm17_method}

The vMF maximum likelihood mean unit vector direction for a sample of random unit vectors (Eq. \ref{e:fisher_3.4}) is simply the normalized sum of the unit vectors in the sample.
Thus, the vMF mean direction, $(\hat{\alpha},\hat{\beta})=(\hat{\theta},\hat{\phi})$, and its accompanying confidence cone (Eq. \ref{e:fisher_5.13}) measure only the random error but not the systematic error in the estimate of the mean pole of the Plutino sample.
When analyzing or modeling observational data, it is important to consider the effect of observational selection biases.
\citet{vm17} showed that any average orbit pole of (non-resonant) KBO samples computed by simply adding up the orbit normal vectors is significantly affected by selection biases in the surveys used to compile the sample population, and often does not reliably reflect the true mean orbit pole of the  population.
Similar observational selection effects would apply to the resonant Kuiper belt objects, including the Plutinos investigated here.

In order to mitigate selection biases and identify a debiased mean pole, we follow the method pioneered by \citet{bp04} and elaborated by \citet{vm17}.
In this method, the mean plane of a Plutino sample is identified as the plane of symmetry of the sky-plane velocity vectors.
Given the barycentric orbital elements of each Plutino, we can compute its barycentric unit position directional vector $\hat{\bm{r}}_i$ and its barycentric unit orbit normal vector $\hat{\bm{h}}_i$.
The sky-plane velocity directional unit vector of each Plutino is $\hat{\bm{v}}_{t,i}=\hat{\bm{h}}_i\times\hat{\bm{r}}_i$.
If a Plutino's orbit normal vector is precisely aligned with the mean pole $\hat{\bm{n}}$ of the population, its sky-plane velocity vector $\hat{\bm{v}}_{t,i}$ will be precisely orthogonal to $\hat{\bm{n}}$, such that $\hat{\bm{n}}\cdot\hat{\bm{v}}_{t,i}=0$.
In a statistical sense, we seek the nonzero value of $\hat{\bm{n}}$ that minimizes
\begin{equation}
J=\sum_{i=1}^n |\hat{\bm{n}}\cdot\hat{\bm{v}}_{t,i}|.
\label{e:vm17_cost_function}
\end{equation}
If we define
\begin{equation}
\hat{\bm{n}}=(\sin{\theta_0}\cos{\phi_0},\sin{\theta_0}\sin{\phi_0},\cos{\theta_0}),
\label{e:define_n_hat}
\end{equation}
then it is convenient to evaluate Eq.~\ref{e:vm17_cost_function} on a grid in $\theta_0$ and $\phi_0$ to find the $\hat{\bm{n}}$ that minimizes $J$.

We compute the confidence limits of $\hat{\bm{n}}$ just as for $\hat{\bm{x}}$, using the spherical standard error as in Eq.~\ref{e:fisher_5.11}--\ref{e:fisher_5.13}, but replacing $\hat{\bm{x}}$ with $\hat{\bm{n}}$ in Eq.~\ref{e:fisher_5.11}.

\section{Sample Selection}
\label{s:Plutinos}

To identify the 3:2 resonant population -- the Plutinos -- for this study, we first used the JPL Solar System Dynamics Group's Small Body Database Query \citep{jpl_sbdb} to retrieve all objects with constraints of heliocentric $38<a<42$ au, $e<1$.
On 2022 October 12, this returned 887 objects.
We eliminated all objects without specified semimajor axis uncertainty or with fractional semimajor axis uncertainty $\sigma_a/a>0.05$, making sure to keep Pluto.

Next, we downloaded the MPCORB.DAT database from the MPC on 2022 October 12 \citep{mpcorb}  and cross-referenced it against the Small Body Database to eliminate all objects that have been observed for fewer than the three oppositions recommended by \citet{gmv08}.
These two steps were meant to exclude objects with orbits too uncertain to be reliably classified.

In the third step, we used the \textsc{Python} package \textsc{Astroquery} \citep{2019AJ....157...98G} to retrieve barycentric elements for each remaining object at 2022 January 1 from JPL Horizons\footnote{\url{https://ssd.jpl.nasa.gov/horizons/app.html}}.
We then discarded all objects outside the region $1.28\,a_\textrm{N}\leq a\leq1.34\,a_\textrm{N}$, where $a_\textrm{N}=30.0$ au is the barycentric semimajor axis of Neptune on 2022 January 1.
These semimajor axis limits are the approximate lowest and highest semimajor axes for the 3:2 MMR as computed by \cite{lm19}.
This left 690 objects.

Finally, for each of the 690 remaining objects, we made plots of the 3:2 critical resonant angle versus time starting from the epoch 2022 January 1 (for the duration of a 10 Myr integration with the $n$-body integrator \textsc{rebound}) and examined each plot by eye.
We also employed the resonance identification algorithm detailed in \citet{yu18} to back up our visual inspection.
This left 431 Plutinos.
This list of Plutinos is available in electronic-readable form in the online version of this paper.
Table \ref{table:plutino_list} provides a sample of the first few lines of this list.

The above method of identifying the Plutinos does not account for orbital uncertainties, which in the Kuiper belt can, in some cases, be large enough to render an object's membership in a particular resonant population insecure.
We examined the effect of orbital uncertainties on the Plutino sample identification by generating clones within the uncertainties and found that neither the sample nor the results of the vMF model of inclination distribution are significantly affected.
Additional details are given in
Appendix~\ref{clones}.

\begin{table}
\caption{This table reports the packed Minor Planet Center identification code and barycentric semimajor axis $a$, eccentricity $e$, inclination $i$, longitude of ascending node $\Omega$, argument of perihelion $\omega$, and mean anomaly $M$ for a few arbitrary Plutinos as of 2022 January 1 (JD 2459580.5).
The table accompanying the online form of this article reports the orbital elements of all 431 Plutinos with the full accuracy given by JPL Horizons via \textsc{Astroquery}.
All elements are in the J2000 reference frame.}
\label{table:plutino_list}
\begin{tabular}{lllllll}
\hline
MPC ID & $a$ (au) & $e$ & $i$ & $\Omega$ & $\omega$ & $M$ \\
\hline
15789 & 39.5 & 0.19 & $5.2^\circ$  & $355^\circ$ & $317^\circ$ & $74^\circ$ \\
15810 & 39.4 & 0.12 & $3.8^\circ$  & $145^\circ$ & $102^\circ$ & $38^\circ$ \\
20108 & 39.5 & 0.15 & $19.6^\circ$ & $188^\circ$ & $143^\circ$ & $68^\circ$  \\
24952 & 39.4 & 0.23 & $16.6^\circ$ & $347^\circ$ & $82^\circ$  & $346^\circ$ \\
47171 & 39.4 & 0.22 & $8.4^\circ$  & $97^\circ$  & $295^\circ$ & $10^\circ$   \\
\hline
\end{tabular}
\end{table}

\section{The vMF Distribution of the Plutinos}
\label{s:plutinos_results}

\begin{figure}
\includegraphics[width=0.9\columnwidth]{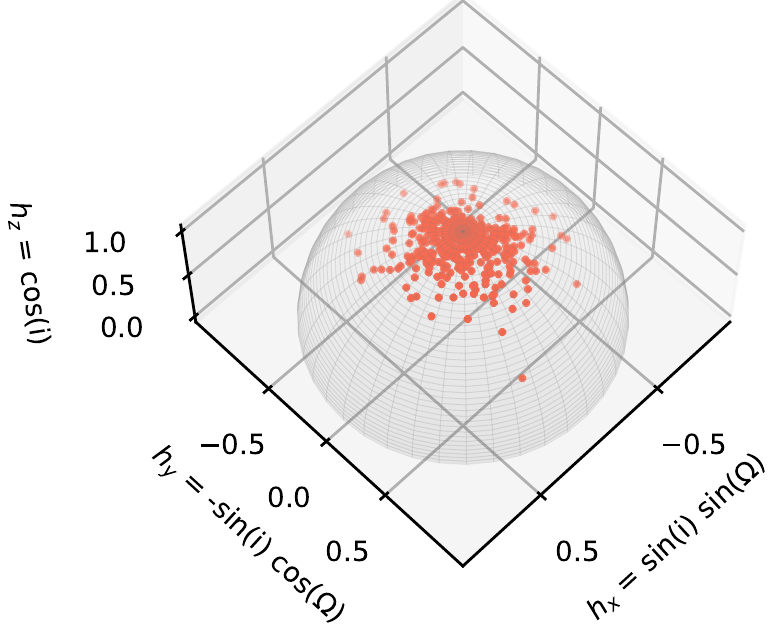}
\caption{The orbit poles of the 431 Plutinos as a scatter plot on the unit sphere (J2000 coordinate system).}
\label{fig:fig_sphere}
\end{figure}

In this section, we report numerical results for the calculations set forth in Sections~\ref{s:vmf_distribution}--\ref{s:vm17_method}.
We apply these calculations to the orbit normal unit vectors $\hat{\bm{h}}_i$ of the Plutinos, substituting $\hat{\bm{h}}_i$ for $\bm{x}_i$ in all equations starting with Eq.~\ref{e:fisher_3.2}.
As shown in Fig. \ref{fig:fig_sphere}, the orbit poles of the 431 Plutinos occupy a portion of the unit sphere with appreciable curvature, inviting the application of directional statistics to describe them.
They can easily be assumed to cluster around a single mean direction, and their longitudinal distribution does not, at first glance, appear too irregular to justify the vMF distribution's assumption of rotational symmetry.
The numbers and figures for the remainder of the paper use barycentric orbital elements for 2022 January 1.

\subsection{Mean direction}
\label{ss:mean_direction_results}

Following the procedure detailed in Section~\ref{ss:estimate_sample_mean}, we find the mean direction of the Plutino samples to be $\hat{\bm{x}}=(0.051,0.035,0.998)$, with colatitude $\hat{\alpha}=3.57^\circ$ and longitude $\hat{\beta}=34.38^\circ$.
This gives the mean plane of the Plutinos an inclination of $i_0=3.57^\circ$ and a longitude of ascending node of $\Omega_0=124.38^\circ$.

\subsection{Mean direction confidence region}
\label{ss:confidence_region_results}

From Section~\ref{ss:estimate_confidence_region}, a 99.7 per cent confidence cone around the mean direction of the Plutino samples has a half-angle of $q=1.68^\circ$.

\subsection{Concentration parameter}
\label{ss:kappa_results}

From Section~\ref{ss:estimate_kappa}, we estimate the concentration parameter of the Plutino samples as $\hat{\kappa}=31.6$.
The 99.7 per cent confidence interval for $\hat{\kappa}$ is $(27.3,\,36.3)$.

\subsection{Rayleigh width parameters}
\label{ss:rayleigh_results}

From Eq.~\ref{e:sigma_i}, we find the relative colatitudes (relative inclinations) of the Plutinos have a Rayleigh width parameter $\sigma\approx10.2^\circ$.
The maximum-likelihood estimate from Eq.~\ref{e:dl_dsigma} is $\hat{\sigma}_{\rm{MLE}}=10.3^\circ$.
Using the relationship of $\sigma$ to $\kappa$ (Eq.~\ref{e:sigma_i}), we transform the 99.7\% confidence interval of $\hat\kappa$ $(27.3,36.3)$, into the corresponding 99.7\% confidence interval of $\hat\sigma$ as $(9.5^\circ,11.0^\circ)$.

\subsection{Debiased mean direction}
\label{ss:vm17method_results}

From Section~\ref{s:vm17_method}, we find a debiased mean pole for the Plutinos of
\begin{equation}
\hat{\bm{n}}=(0.0152,-0.0364,0.9992)
\end{equation}
with $\theta_0=2.26^\circ$ and $\phi_0=292.69^\circ$.
The debiased mean pole has an inclination of $i_0=2.26^\circ$ and a longitude of ascending node of $\Omega_0=22.69^\circ$.
The 99.7 per cent confidence cone around the debiased mean pole has a half-angle of $q=1.69^\circ$.

\section{Discussion}
\label{s:discussion}

\begin{figure}
\includegraphics[width=0.9\columnwidth]{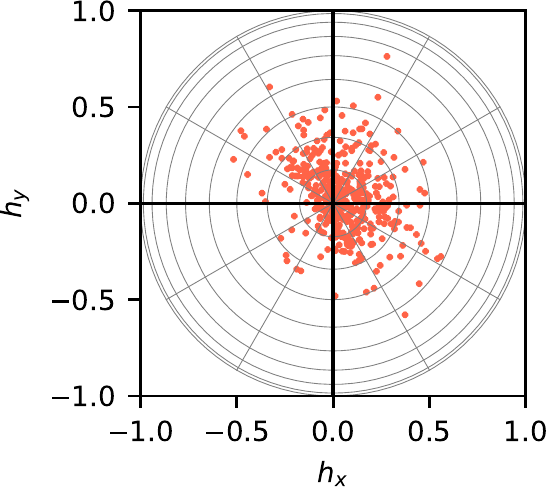}
\caption{Plutino orbit poles projected in the ecliptic plane. Longitude lines are drawn at $30^\circ$ intervals and latitude circles are drawn at $10^\circ$ intervals from the ecliptic pole to the equator.  (Note that an orbit pole is located at a ecliptic longitude angle $\Omega-90^\circ$ because the projection of an orbit pole vector in the ecliptic is $+90^\circ$ away from the longitude of ascending node, $\Omega$, of the orbit plane on the ecliptic; recall that $(h_x,h_y)=\sin{i}\,(\sin{\Omega},-\cos{\Omega})$.)
}
\label{fig:fig_hxhy}
\end{figure}

\begin{figure}
\includegraphics[width=0.9\columnwidth]{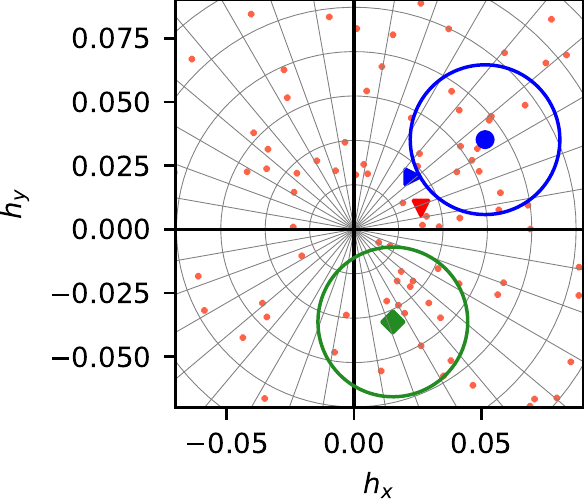}
\caption{Detail of Fig. \ref{fig:fig_hxhy} near the origin. Longitude lines are drawn at $10^\circ$ intervals and latitude circles are drawn at $1^\circ$ intervals starting from the ecliptic pole at the origin. Orbit poles of individual Plutinos are indicated with the small red dots. The vMF mean pole is the blue point, surrounded by a blue 99.7 per cent confidence cone. The debiased mean pole is the green diamond, surrounded by a green 99.7 per cent confidence cone. The invariable pole of the solar system is the red $\bigtriangledown$. The orbit pole of Neptune is the blue $\rhd$.
}
\label{fig:fig_hxhy_detail}
\end{figure}

In Fig.~\ref{fig:fig_hxhy}, we plot the orbit poles of all 431 Plutinos projected in the ecliptic plane; that is, we plot $(h_x,h_y)$ for the entire Plutino sample.
In Fig.~\ref{fig:fig_hxhy_detail}, we plot a detail view near the origin.
The detail view shows the mean pole of the Plutinos, $\hat{\bm{x}}$, and its 99.7 per cent confidence circle; the debiased mean pole $\hat{\bm{n}}$ and its 99.7 per cent confidence circle; and the orbit pole of Neptune and the pole of the invariable plane of the Solar System.
As Neptune is the dominant perturbing planet for the Plutinos, and Neptune's orbit pole itself precesses on secular timescales under the influence of the other planets, it is reasonable to expect that the Plutinos' mean pole would bear some relationship to Neptune's pole and/or to the invariable pole.
Our results show that the invariable pole of the Solar System is separated from $\hat{\bm{x}}$ by $2.1^\circ$ and from $\hat{\bm{n}}$ by $2.6^\circ$.
The orbit pole of Neptune is separated from $\hat{\bm{x}}$ by $1.8^\circ$ and from $\hat{\bm{n}}$ by $3.3^\circ$.
The two Plutino mean plane estimates $\hat{\bm{x}}$ and $\hat{\bm{n}}$ are separated by $4.6^\circ$, which is great enough that their confidence circles do not overlap; thus, the two estimates contradict each other at the 99.7 per cent level.
Neither confidence circle contains the invariable plane or the orbit pole of Neptune.
Because the two mean plane estimates are so widely separated, we can draw no conclusions about any dynamical relationship between the mean plane of the Plutinos and the orbit pole of Neptune or the invariable plane.

\begin{figure}
\includegraphics[width=2.5in]{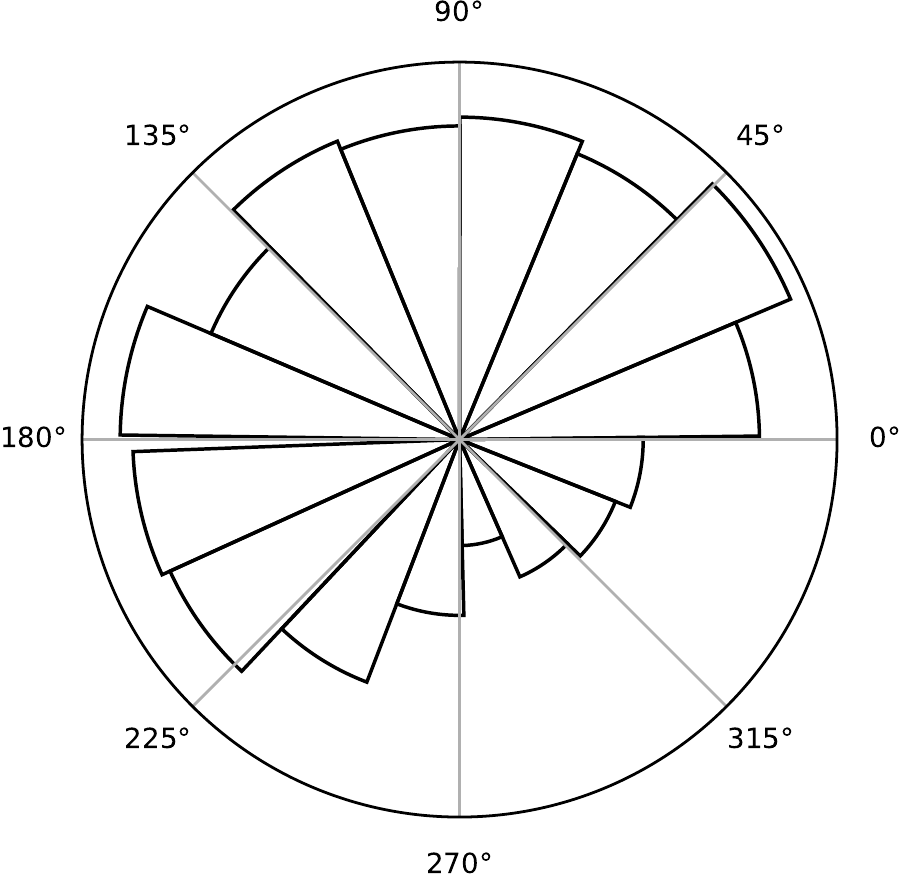}
\caption{Area-weighted polar histogram of $\Omega$, the longitudes of ascending node relative to the ecliptic plane.
(The area of a sector in a polar histogram corresponds to the height of a bar in a rectangular histogram.)
}
\label{fig:fig_longitudes}
\end{figure}

We observe that the ecliptic longitudes of the Plutino poles are not uniformly distributed.
This is shown more clearly in Fig.~\ref{fig:fig_longitudes} as a circular histogram of $\Omega_i$, the longitudes of ascending node of the Plutinos relative to the ecliptic plane.
The strong non-uniformity of $\Omega_i$ is evidence of selection biases in observational surveys of the outer Solar System, none of which have all-sky coverage and all of which have non-uniform coverage of the near-ecliptic sky \citep[cf.][]{gladman2012resonant,Shankman:2017}.

\begin{figure}
\includegraphics[width=0.95\columnwidth]{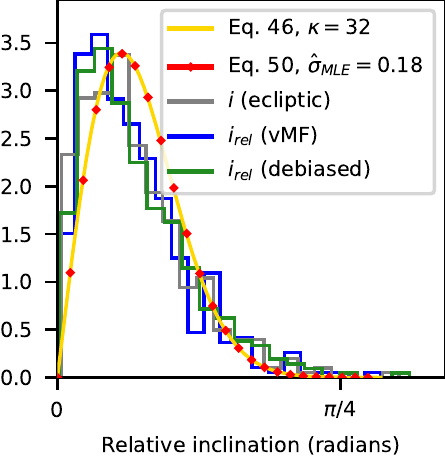}
\caption{Histogram of the Plutino orbit inclinations relative to the ecliptic pole (gray), the vMF mean pole (blue), and the debiased mean pole (green).
The vMF relative inclination pdf (Eq. \ref{e:fisher_4.29b}) with  $\kappa=32\,$ is shown as the yellow curve.
A truncated Rayleigh pdf (Eq. \ref{e:rayleigh2}) with a maximum-likelihood width parameter of $\hat{\sigma}_\textrm{MLE}=10.2^\circ$ (Eq.~\ref{e:dl_dsigma}) is shown as the red dotted curve.
}
\label{fig:fig_inclinations}
\end{figure}

In Fig. \ref{fig:fig_inclinations}, we plot histograms of the inclinations of the Plutinos relative to the ecliptic plane (gray), the vMF mean pole (blue), and the debiased mean pole (green).
The one-dimensional vMF inclination distribution (Eq.~\ref{e:fisher_4.29b}) with $\kappa=32$ is plotted in yellow, and the maximum-likelihood truncated Rayleigh pdf (Eq.~\ref{e:rayleigh2}--\ref{e:dl_dsigma}) is plotted in red.
The vMF concentration parameter of the Plutinos is large enough that the maximum-likelihood truncated Rayleigh approximation is nearly identical to the vMF distribution.
The inclination histograms of the Plutinos relative to the vMF mean pole and the debiased mean pole are nearly identical, and are skewed towards slightly lower inclinations than the histogram relative to the ecliptic pole.
Visual inspection shows that the data have greater positive skewness than the fitted vMF and maximum-likelihood truncated Rayleigh inclination distributions.
They are more highly concentrated at low inclinations, yet they have a longer high-inclination tail.

Even though the inclination distribution of the Kuiper belt observational sample is noticeably skewed relative to the one-dimensional vMF inclination distribution and the Rayleigh distribution, Gaussian and Rayleigh-like distributions have been widely used and continue to be useful for modeling the orbit plane distributions of Kuiper belt populations.
Observational incompleteness of samples and observational selection biases are likely contributing factors for the poor fits.
We anticipate that the vMF distribution can also be useful in modeling observational samples, despite present shortcomings. In the case of synthetic or simulated data, the vMF distribution can be a natural fiducial model for the distribution of orbital planes.
We note that many two-parameter probability distributions with support on the positive real line could potentially be fit to the relative inclination distribution of the Plutinos with high accuracy, but, lacking a physical motivation for any particular two-parameter distribution, we do not explore that direction at this time.

\subsection{Comparison to past studies}
\label{ss:comparison_to_past_studies}

We report a vMF concentration parameter for the Plutinos of $\hat{\kappa}=31.6$, with a 99.7 per cent confidence interval of  $(27.3,36.3)$ and a 68 per cent confidence interval of $(30.1,33.1)$.
This corresponds to a truncated Rayleigh width parameter of $\sigma=10.2^\circ$, with a 99.7 per cent confidence interval of $\sigma=10.2^{+0.8}_{-0.7}$ degrees and a 16th-84th percentile confidence interval of $\sigma=10.2^{+0.2}_{-0.4}$ degrees.

\citet{brown_2001, gladman2012resonant, ossos3} investigated the orbital plane distributions of the Plutinos but did not report a mean pole measurement.
These authors fitted the Plutinos' inclinations relative to the ecliptic to the one-parameter functional form of Eq.~\ref{e:fiB01} and reported the best-fitting width parameter $\sigma$.  \citet{brown_2001} reported the 16th-84th percentile confidence interval for $\sigma$ of the Plutinos to be $\sigma=10.2^{+2.5}_{-1.8}$ degrees, \citet{gulbis2010unbiased} gave the same interval to be $\sigma=10.7^{+2.0}_{-2.3}$ degrees, \citet{gladman2012resonant} reported $\sigma=16^{+8}_{-4}$ degrees, and \citet{ossos3} reported  $\sigma=12^{+1}_{-2}$ degrees.

Our Plutino sample has a generally similar but smaller inclination dispersion (as well as smaller uncertainty in the dispersion measure) compared with any previous study but \citet{brown_2001}.
Our smaller measured dispersion may be partly owed to the fact that we measured the inclination dispersion relative to the measured mean pole of the sample, distinct from the ecliptic pole, and partly also to the larger size of our sample.
We have 431 Plutinos, whereas \citet{brown_2001} had a sample of 20, \citet{gulbis2010unbiased} had a sample of 51, \citet{gladman2012resonant} had a sample of 24, and \citet{ossos3} had a sample of 21.
\citet{gulbis2010unbiased}, \citet{gladman2012resonant} and \citet{ossos3} restricted their Plutino samples to detections from the well-characterized DES, CFEPS and OSSOS surveys, whereas we use the entire catalog of known KBOs.

\subsection{Directions for future work}
\label{ss:future_work}

In this paper, we compute the mean plane and its uncertainty region for the observed Plutino sample using two methods, the vMF mean pole $\hat{\bm{x}}$ and the debiased mean pole $\hat{\bm{n}}$, but we make no attempt to explain the location of the mean plane of the Plutinos.
In future work, we hope to apply theory and simulation to find where the mean plane of the Plutinos ``should'' be as forced by the known planetary perturbers, in order to comprehend the significance of these results.
We also hope to extend the vMF distribution, or vMF mixture distributions, to other Kuiper belt populations, including the cold classical KBOs and the entire Kuiper belt taken as a whole.
It would also be useful to undertake a detailed investigation of the reliability of the plane-of-symmetry mean pole $\hat{\bm{n}}$ in identifying the true mean pole as compared to $\hat{\bm{x}}$ (from the raw average of the orbit poles), under a variety of survey designs, or otherwise harmonize the two estimates.
Other possible directions for future work include developing a method to compute a more rigorous model-independent estimate of the uncertainty in $\hat{\bm{n}}$ and deriving constraints on unseen perturbers from the deviation of the measured mean pole from the theoretically expected one based on known perturbers.

\section{Summary}
\label{s:summary}

In this paper, we proposed the two-dimensional von Mises--Fisher distribution on the surface of a unit sphere as a physically motivated functional form for the distribution of orbit poles of populations of small bodies in the solar system.
We derived its relationship to the univariate Rayleigh distribution (for  inclinations) and to the Gaussian distribution (for the components of the two-component inclination vector) that are more commonly used to model inclination distributions of small bodies.

We applied the vMF distribution to the current observational sample of Plutinos, and obtained the following results.

\begin{enumerate}
    \item We found that the Plutinos' vMF mean pole is given by the unit vector $\hat{\bm{x}}=(0.051,0.035,0.998)$, with a 99.7 per cent confidence cone of half-angle $1.68^\circ$ (in the J2000 coordinate system).
    The inclination and longitude of ascending node of the vMF mean pole are $3.57^\circ$ and $124.38^\circ$, respectively.

    \item  We found the vMF concentration parameter of the Plutinos as $\hat{\kappa}=31.6$, with 99.7 per cent confidence bounds of $(27.3,36.3)$.
    This resembles a Rayleigh distribution function (for the inclination relative to the mean pole) of width parameter $\sigma\approx10.2^\circ$ and 99.7 per cent confidence bounds of $(9.5^\circ,11.0^\circ)$.

    \item We computed the debiased mean pole of the Plutinos as the pole of the plane of symmetry of their sky-plane velocity vectors. This is found to be $\hat{\bm{n}}=(0.0152,-0.0364,0.9992)$, with a 99.7 per cent confidence cone of half-angle $1.69^\circ$.
    The inclination and longitude of ascending node of the debiased mean pole are $2.26^\circ$ and $22.69^\circ$, respectively.

    \item The vMF mean pole and debiased mean pole of the Plutinos are widely separated, by $\sim4.6^\circ$.
    Their 99.7 per cent confidence cones do not overlap.
    Neither confidence cone contains the invariable pole of the Solar System or the orbit pole of Neptune.

\end{enumerate}

Our Plutino sample has a smaller inclination dispersion and smaller uncertainty than any previous study but \citet{brown_2001}, possibly because our sample size is much larger.
The Plutino sample is more highly concentrated at low inclinations and has a heavier high-inclination tail than the fitted vMF distribution.
Nevertheless, we assert that the vMF distribution function is a natural functional form to adopt in theoretical models and in simulated or observational data of random directional vectors.

\section*{Acknowledgements}

ICM performed the computations, with advice from RM and JTK. RM designed the research, and RM and ICM collaborated on the writing of the paper.
The authors acknowledge funding from the JPL SURP and SIP student support programs and from the NSF (grant AST-1824869).
A portion of this research was carried out at the Jet Propulsion Laboratory, California Institute of Technology, under a contract with the National Aeronautics and Space Administration (80NM0018D0004).
RM thanks the Canadian Institute for Theoretical Astrophysics for hosting a sabbatical visit during the late stages of this work.
We thank the editorial staff and the reviewer Matija Cuk for their comments, questions, and advice.

\section*{Data Availability}

This research has made use of data and/or services provided by the International Astronomical Union's Minor Planet Center.
No new data were generated in this work.
Table \ref{table:plutino_list}, containing the Plutino sample used in this work, can be found in full with the online edition of this article.
All of the code pertaining to this paper can be downloaded from GitHub at \url{https://github.com/iwygh/mmk23a}.
For the MPC and JPL datasets of 2022 October 12, contact the authors.

\bibliographystyle{mnras}
\bibliography{refs_for_arxiv}

\appendix

\section{Accounting for Orbital Uncertainties}\label{clones}

Orbital uncertainties in the Kuiper belt can render an object's membership in a particular resonant population insecure.
To see whether our Plutino sample would change when accounting for orbital uncertainties, we used a separate resonant object identification pipeline that began with the same 690-object semimajor axis limits as in Section \ref{s:Plutinos}.
Our procedure for securely identifying Plutinos does not, like the laborious standard method described in \citet{gmv08} and used in \citet{vm17} and \citet{sv20}, have the advantage of returning to the observations and fitting new orbits from first principles, but it is more portable and more highly automated.
We begin by building off the work of \citet{sv20}.

\citet{sv20} used the system of \citet{gmv08} to classify 2305 KBOs from the Minor Planet Center (MPC) database as of 2016 October 20.
They then used the \textsc{Python} machine learning package \textsc{Scikit-Learn} \citep{scikit-learn} to develop a gradient boosting classifier for fast, easy classification of KBOs as either Classical, Scattering, Detached, or Resonant, training said classifier upon the orbits newly classified using the \citet{gmv08} criteria.
Their code integrates a KBO orbit from initial barycentric elements in the \textit{n}-body integrator \textsc{rebound} \citep{rebound} for 100 kyr and records 55 features of the orbit for use by the machine learning algorithm.
Full details of the settings used for the machine learning algorithm, and a full explanation and justification of the 55 recorded features, are given in their paper.
To allow other researchers to use their gradient boosting classifier to classify KBOs from their barycentric elements without repeating the entire process of selecting and justifying the features of the orbit to record and the settings to use for the algorithm, \citet{sv20} posted user-friendly sample code and training data to GitHub.

To classify the sample of 690 objects remaining after the first three steps of the downselection process in Section \ref{s:Plutinos}, we downloaded the \textsc{Python} sample code and training data (KBO\_features.csv) from the \citet{sv20} GitHub repository.
We used their gradient boosting classifier without modification and trained it on the same training set they provided, exactly as suggested in the sample code.

To account for orbital uncertainties, we used the JPL Small Body Database API \citep{jpl_sbdb_api} to download a JSON file for each of the 690 objects.
The JSON file contained a nominal heliocentric orbital state, a 6x6 covariance matrix for the heliocentric orbit, and an epoch for the nominal orbit and the covariance matrix.
The heliocentric orbital elements and their covariance were given as $e,q,t_p,\Omega,\omega,i$, i.e. eccentricity, perihelion distance in au, time of perihelion passage (Julian date), longitude of the ascending node, argument of perihelion, and inclination, where all angles are in degrees and referenced to the J2000 plane, and the epoch is a Julian date.

We generated 301 heliocentric orbital element sets for each object: the nominal orbit, and 300 clones from a Gaussian distribution centred at the nominal orbit, from the given covariance.
The mean anomaly for each orbital element set was computed as the mean motion for the semimajor axis, times the elapsed time between the time of perihelion passage and the epoch.
We used Horizons to download heliocentric orbital elements for the giant planets at each epoch.
For each of 690 objects, we then built 301 \textsc{rebound} simulations consisting of the outer planets at the appropriate epoch and the orbital element set of the nominal orbit or the clone at the same epoch.

Each \textsc{rebound} simulation was then classified as Classical, Scattering, Detached, or Resonant using the unmodified \citet{sv20} gradient boosting classifier.
If all 301 of the 301 orbital element sets for each object were classified as Resonant, we kept the object; otherwise, we discarded it.
After this step, the 690 objects were reduced to 427.

The preceding steps were meant to ensure that only objects securely in mean motion resonances (MMRs) remained in our sample, but did not guarantee that the Resonant objects would librate in the 3:2 MMR.
To identify which Resonant objects were truly Plutinos, we took the first orbital element set identified as Resonant for each of the 427 Resonant objects (and the corresponding orbital elements of the giant planets at the same epoch), and integrated for 10 Myr, recording the 3:2 angle every 100 yr.
We employed the resonance identification algorithm detailed in \citet{yu18} to classify the time history of the 3:2 angle for each of the 427 objects as librating (True) or circulating (False).

At the conclusion of this Plutino identification pipeline, we had a slightly different sample than we previously found without accounting for orbital uncertainties.
In Section \ref{s:Plutinos}, we found 431 Plutinos, but in this section, we found only 424.
We repeated all calculations for the 424-object sample, and the results were negligibly different from those of the calculations that used the 431-object sample.

\bsp
\label{lastpage}
\end{document}